\documentclass{elsart}
\usepackage[dvips]{graphicx}
\usepackage{amssymb} 
\usepackage{bm} 
\journal{Chemical Physics Letters}
\begin{document}
\begin{frontmatter}
\title{Generation of multiple circular walls on a thin film of nematic liquid crystal by laser  scanning}

\author{M. Kojima},
\author{J. Yamamoto},
\author{K. Sadakane},  
\author{K. Yoshikawa\corauthref{cor}}
\corauth[cor]{Corresponding author.}  
 \ead{yoshikaw@scphys.kyoto-u.ac.jp} 
\address[kyoto]{Department of Physics, Graduate School of Science, Kyoto University \& Spatio-temporal Project, ICORP, JST, Kyoto 606-8502,  Japan\\}

Received 11 February 2008; in final form 2 Marc 2008h

\begin{abstract}
We found that multiple circular walls (MCW) can be generated on a thin film of a nematic liquid crystal through a  
 spiral scanning of a focused IR laser.
The ratios between the radii of adjacent rings of MCW were almost constant.
These constant ratios can be explained theoretically by minimization of the Frank elastic free energy of nematic medium.
The director field on a MCW exhibits chiral symmetry-breaking although the elastic free energies of both chiral MCWs are degenerated, i.e., 
the director on a MCW can rotate clockwise or counterclockwise along the radial direction.
\end{abstract}
\end{frontmatter}

\section{Introduction}
Liquid  crystal (LC) phases bring about a rich variety of textures as visualized by polarizing microscopy \cite{Dierking}.
This polymorphism in the texture results from the inhomogeneity in molecular alignment due to  defects in the LC \cite{deGennes}.
Over the past few decades, much attention has been paid to colloidal dispersions in nematic LC 
because of their peculiar properties, such as a  topological defect around a particle and long-range interaction among the particles \cite{Poulin_Science,Poulin_PRE,Stark}.
This long-range interaction also results from inhomogeneity in the molecular alignment. 
The averaged molecular alignment is expressed as the director \cite{deGennes}.
It has been found that a strong laser beam induces a distortion in the director field of nematic LC \cite{Durbin}.
Especially, a linearly polarized laser can orient the director in the illuminated region in the direction of laser polarization \cite{Gibbons,Simons}.
It has been reported that defects in LC can be modified by use of laser manipulation \cite{Janossy,Hotta}.
Recently, M{\v u}sevic et al. combined the property of a defect around a colloidal particle suspended in nematic LC 
with distortion in the nematic LC induced by a laser beam. 
They found that a particle with a lower refractive index than that of the surrounding nematic medium 
can be picked up with optical tweezers \cite{Musevic},
although such a particle  cannot be trapped in isotropic media.
{\v S}karabot et al. showed theoretically that such extraordinary trapping is achieved through the interaction 
between the laser-induced distortion in the director field and a topological defect near the particle \cite{Skarabot}.
Thus, the interaction between a defect in LC  and a laser beam produces various unique phenomena. 
Here, we tried  to generate a new pattern in a nematic LC by using the interaction between a defect and a laser beam.

\section{Materials and methods}
The nematic material 5CB (Tokyo Chemical Industry co., Japan) was put into a microtube. 
Pure water was dispersed in the microtube. 
The nematic containing water droplets was vortexed.  
The microtube was then centrifuged for a short period to eliminate large water droplets. 
The nematic containing micron-sized droplets was placed between glass slides with a thickness a few $\rm\mu m$. 
The thickness was estimated by dividing the volume of nematic by the contact area on the glass surface.
The nematic was sheared to easily give a Schlieren texture, and the glass slides were baked at 500 $^\circ$C for an hour before use.

Observations were performed through a polarizing microscope (converted IX70, Olympus, Japan) 
equipped with a $\times$ 100 oil immersion objective lens (UPlan Apo IR, N.A. 1.35, W.D. 0.1 $\rm mm$, Olympus, Japan).
A linearly polarized Nd:YAG laser with a focus of $\sim$ 1 micrometer was introduced to the nematic by a dichroic mirror 
and the objective lens. 
The linearly polarized laser was constructed with a randomly polarized laser source (JOL-D8PK-Y, JENOPTIK, Germany) and a polarization beam splitter.
The direction of laser polarization was controlled by a half-wave plate.
The laser emission power was 2.0 W, as calibrated just before the objective lens.
The observation and laser irradiation area in the nematic were controlled with a motorized stage (BIOS-302T, Sigma-koki, Japan).

\section{Results}
\begin{figure*}
\includegraphics[scale=0.8]{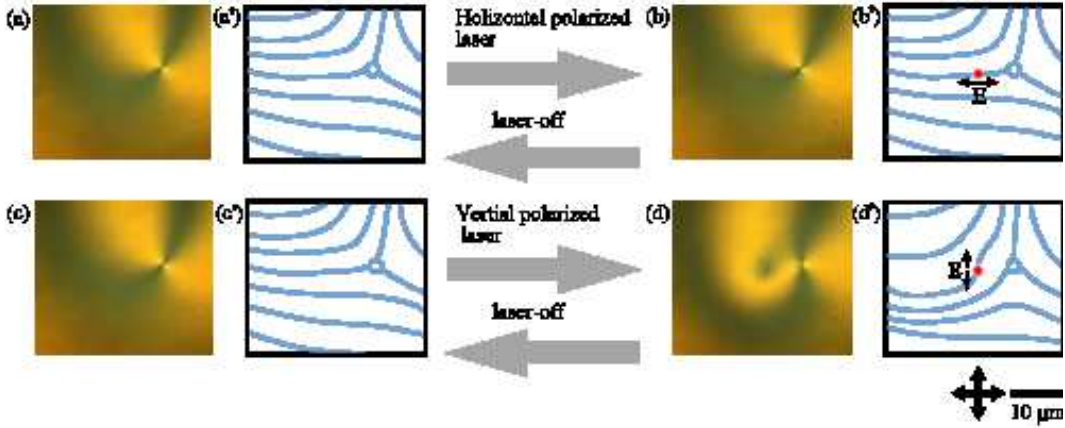}
\caption{
(color online). 
Responses of a brush on a thin film of liquid crystal for polarized laser beams.
(a), (c) A wedge disclination with a strength of -1/2, just before laser irradiation. 
(b) A horizontally polarized laser beam illuminates the brush.
(d) A vertically polarized laser illuminates the same region as in (b).  
The brush is only repelled from the vertically polarized laser spot.
In both cases, the textures recovered almost reversibly when the laser irradiation was shut off.
Figures (a')-(d') are schematic illustrations of the director fields of (a)-(d) respectively.
The red circles in the primed figures represent the position of the beam spot. 
The double-headed arrow represents the direction of the polarizer and analyzer in polarizing microscopy.
\label{steady}}
\end{figure*}

Figure \ref{steady} shows the responses of a brush distributed from a wedge disclination for horizontally and vertically polarized laser beams.
The disclination was pinned to the glass substrate by chance.
The strength of the disclination was -1/2, 
as judged from the response in the texture for a simultaneous rotation of the polarizer and analyzer of the polarizing microscopy while maintaining crossed nicols.
When the brush was illuminated by a horizontally polarized laser beam (Fig. \ref{steady} (a)), the texture showed a minute change  (Fig. \ref{steady} (b)). 
On the other hand, with a vertically polarized laser beam, the illuminated brush was repelled from the beam spot and the texture was  
completely changed from Fig. \ref{steady} (c) to Fig. \ref{steady} (d).
The responses of the neighboring brush connected to an identical disclination core for polarized beams were inverted: 
the neighboring brush was only repelled from the horizontally polarized laser beam.
When the laser was shut off, the conformation of the brush returned back to the texture seen before the laser irradiation. 
The distances between the repelled brush and laser spot were distributed broadly.  
We confirmed experimentally that brushes which grew from disclination cores ($\pm$ 1/2 and $\pm$ 1) are repelled from either a horizontally or vertically polarized beam spot. 
The responses of a disclination core to a linearly polarized laser beam have been reported by Hotta et al. \cite{Hotta}.

\begin{figure}
\includegraphics[scale=0.8]{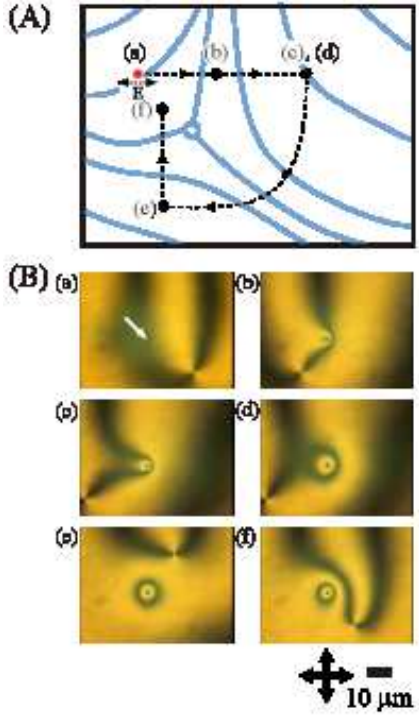}
\caption{ 
(color online). 
Generation of a single-ring wall by laser scanning.
(A) Schematic representation of the trajectory of laser scanning around a disclination.
(B) Snapshots at the points labeled in (A).
(a) The laser beam illuminates the region near the brush, where the director of the LC is parallel to the laser polarization. 
(b) The laser beam repels the brush, where the director is perpendicular to the laser polarization. 
(c) The beam spot bulldozes out part of the brush.  
(d) The plucked brush is closed spontaneously and a ring pattern emerges.
(e) The ring is a complete circle  while the beam spot is at rest.
(f) The laser spot covered by a circular wall repels the brush, although the director on the brush is parallel to the laser polarization. 
When the laser beam is shut off, the ring pattern disappears immediately. 
The white arrow in (B)(a) represents the position of the beam spot in the observation area.The position is common in (a)-(f).
Other symbols are identical to those in Fig. \ref{steady}}
\label{dynamical}
\end{figure}

Figure \ref{dynamical} shows the emergence of a single-ring pattern. 
The real time movie of the process is available from the internet \cite{movie1}.
The single-ring pattern centered on the beam spot was generated by a laser scanning along the trajectory depicted schematically in Fig. \ref{dynamical} (A). 
%
%
We have chosen the velocity of the laser scan so as to grasp the brush in a steady manner.
Actually, the spot was moved step-by-step, where typical one step is $\sim$ 20 $\rm\mu m$, and during the step-motion the speed was chosen $\sim$ 50 $\rm\mu m$/s.
The present disclination core is identical to that in Fig. \ref{steady}. Figure \ref{dynamical} (B) (a)-(f) show snapshots 
at each point on the scanning trajectory in Fig. \ref{dynamical} (A).
The beam spot passed through without changing the conformation of the brush (Fig. \ref{dynamical} (a)), 
whereas the neighboring brush was repelled from the beam spot(Fig. \ref{dynamical} (b)).
When the spot was further broken into the brush, 
the plucked part of the brush spontaneously closed and a dark ring centered on the beam spot was formed(Fig. \ref{dynamical} (c)-(d)).
The dark ring kept up with the motion of the laser spot with deformation from  a complete circle.
When the beam spot came to rest, the dark ring relaxed into a symmetrical circular form (Fig. \ref{dynamical} (e)).
The laser spot coated with the dark ring repelled the brush (Fig. \ref{dynamical} (f)), 
which was not repelled in the case of a bare laser spot (Fig. \ref{dynamical} (a)).
We have succeeded in generating this pattern only from $\pm$ 1/2 disclinations. 
When the laser irradiation is shut off, the generated rings disappear within the order of 0.1 s.
%
%
%
The ring pattern could not be generated when the velocity of the laser spot was above 100 $\rm\mu m$/s.
In addition, the generated ring patterns broke down when the velocity of the laser spot was above 100 $\rm\mu m$/s.  
The sizes of the dark rings have a broad distribution. 

\begin{figure*}
\includegraphics[scale=0.8]{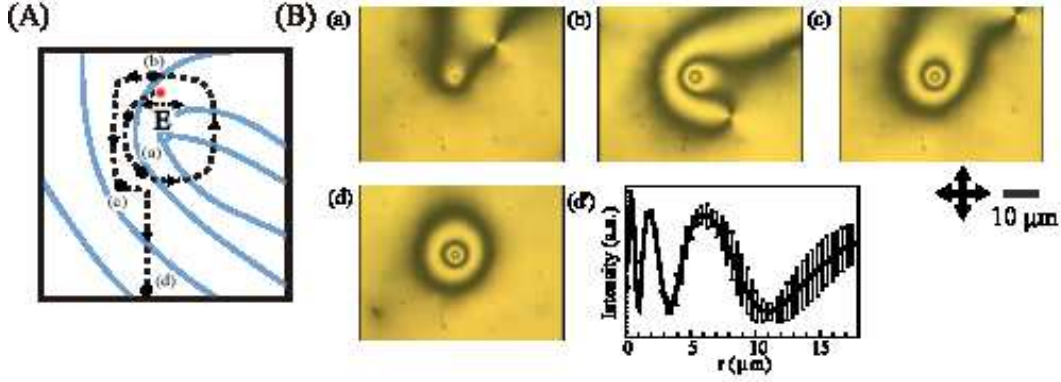}
\caption{
(color online).  
Formation of triple circular walls induced by laser  scanning along the trajectory illustrated in (A) around a +1/2 wedge disclination.
The corresponding snapshots at each labeled point are given in (B). 
(a) Single, (b) double and (c) triple rings appear around the beam spot. 
(d) Time-averaged texture of the triple circular walls over 6 s.
(d') Averaged radial intensity profile of (d) with error bars.
When the laser irradiation is shut off, the rings disappear within a few s.   
The symbols are the same as in the previous figures.}
\label{multi_ring}
\end{figure*}

\begin{figure}
\includegraphics{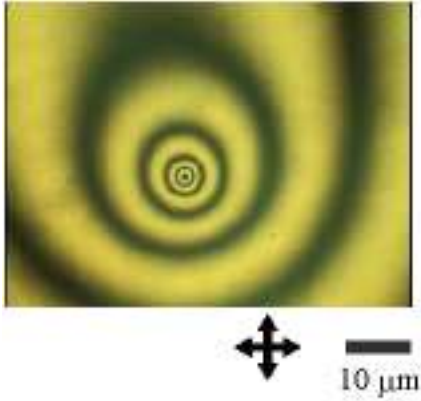}
\caption{
(color online). Quadruple walls generated from a plural number of disclinations through a laser scanning. }
\label{quad}
\end{figure}

Figure \ref{multi_ring} shows the appearance of a triple-ring pattern when the laser spot follows a trajectory $\sim$ a $+$ 1/2 disclination core.
The real time movie of the process is available from the internet \cite{movie2}.
The spot was moved step-by-step, where the typical one step is $\sim$ 20 $\rm\mu m$, and during the step-motion the speed was chosen below 30 $\rm\mu m$/s.
The scanning process corresponds to three iterations of the pattern used to generate a single-ring pattern.
The number of rings increased when the beam spot passed across the brushes (Fig. \ref{multi_ring} (B) (a)-(d)).
The patterns typically measure dozens of micrometers, which is more than 10-fold larger than the size of the beam spot.
When the laser irradiation is shut off, the patterns shrink toward the center. 
The extinction time of the pattern is on the order of a second, which is much longer than that of a single-ring pattern. 
In Fig. \ref{quad} is shown the example of a quadruple-ring, which was generated with a similar procedure. 

In a generating process of the multiple-ring pattern,
a new ring is formed outside of the existing multiple-ring pattern.
When the new ring is created, the inner rings shrink in size.
The radii of dark rings have a broad size distribution. 
The extinction time of a multiple-ring pattern due to shutting-off of the laser is extended with an increase in the number of rings.  

%
%

\section{Discussion}
Let us discuss the mechanism of the change  in the conformation of the brush in Figs. \ref{steady} and \ref{dynamical}.
Since the nematic had a planar configuration, we consider the system to be two-dimensional. 
The interaction between the local director and the laser beam can be interpreted in terms of the change in the dielectric free energy $\Delta F_E$, 
which can be written as \cite{Simons}  
\begin{equation}
\Delta F_E =\int {\rm d}\bm{r} (- \frac{\Delta \epsilon}{4 \pi} |\bm {n}(\bm{r}) \cdot  \bm{E}(\bm{r})|^2)
\end{equation}
where $\Delta \epsilon = \epsilon_\parallel - \epsilon_\perp$ is the anisotropy in the dielectric constants between parallel $\epsilon_\parallel$ and  
perpendicular $\epsilon_\perp$ to the director.
The symbols $\bm{n}(\bm{r})$ and $\bm{E}(\bm{r})$ represent 
the director and oscillating electric field, respectively, of the incident laser beam  at position $\bm{r}$.
Since the nematic phase of 5CB has a positive value in 
$\Delta \epsilon$ ($\fallingdotseq$ 12 $\epsilon_0$, where $\epsilon_0$ is the dielectric constant of vacuum \cite{dielectric} ), 
the director in the illuminated region prefers to be oriented along the direction of laser polarization.

In our experiments, the optical torque for the director is strong enough to completely orient the illuminated director along the direction of laser polarization,
because the beam spot appeared as black circles for both vertically and horizontally polarized laser beams.
As a consequence, the illuminated director behaves as a boundary condition for the director field.
When the illuminated director and polarization of the incident laser beam are parallel, the texture remains constant,
because the illuminated director is already suited for the laser-induced boundary condition.
When the director and laser polarization are perpendicular, the illuminated director is forced to rotate along the direction of polarization of 
the incident laser.
As a result, the director field and texture change so as to satisfy the laser-induced boundary condition.
We noted that the distances between the repelled brush and the laser spot are not constant in the experiments.
This suggests that the interaction between the repelled brush and the laser spot is affected by the laser polarization 
and director distortion due to other disclination cores, distributed in the outside area of the photographs.

Figure \ref{multi_ring} (d') shows the intensity profile of  Fig. \ref{multi_ring} (d).
The radii $r_i$, which  have maximal values in the intensity profile, are approximately $r_1$ = 0.5 $\rm\mu m$, $r_2$ = 1.8 $\rm\mu m$ and $r_3$ = 6.0 $\rm\mu m$, 
where the subscript indicates $i$ = 1, 2, 3. 
In the same way, the radii $r'_i$, which take minimal values, are approximately $r'_1$ = 1.0 $\rm\mu m$, $r'_2$ = 3.4 $\rm\mu m$ and $r'_3$ = 11.0$\rm\mu m$.
Thus, the ratios $r_i/r'_i$ are found to be $\sim$ 1/2. We have confirmed through the experiments that 
this ratio is constant for each triple-ring pattern although the radii of the rings exhibit dispersion. 
We calculate the elastic free energy of the director distortion to explain this trend in the triple-ring patterns.
For simplicity, we assume that the director field on a multiple-ring pattern has cylindrical symmetry on a two-dimensional plane.
Thus, the director $\bm{n}$ on a multiple-ring pattern depends only on the distance $r$ from the center of the beam spot.
Therefore, the director field on a multiple-ring pattern is expressed as ${\bm n}=(n_x, n_y, n_z)= (\cos \psi(r), \sin\psi(r), 0)$, 
where $\psi$ is the azimuthal angle of the director.
With one constant approximation of the elastic constants \cite{deGennes}, 
the elastic free energy $F_{\rm ela}$ can be written as
\begin{equation}
F_{\rm ela}=\int \frac{1}{2} K(\partial_\alpha n_\beta\partial_\alpha n_\beta){\rm d}\bm{r}.
\end{equation}
where $K$ is the elastic constant of nematic medium.
We consider the boundary conditions of a multiple-ring pattern as follows.
Since the director on the beam spot is adjusted to  the direction of laser polarization in the experiments,
we adopt $\psi(r_{\rm c})=\psi_{\rm c}$ as an inner boundary condition, where $r_{\rm c}$ corresponds to the radius of the beam spot 
($2r_{\rm c}\simeq$ wavelength $\lambda$, $\sim$ 1 $\rm\mu m$) and $\psi_{\rm c}$ is the azimuth angle of the direction of laser polarization. 
The parameter $r_{\rm b}$ is introduced as a cut-off length at which the director recovers the orientation angle $\psi_{\rm b}$ in the bulk, where $r_{\rm b}$ is several tens of $\rm\mu m$. 
If we minimize the elastic free energy,  $\psi(r)$ is given as 
\begin{equation}
\label{director_sol} 
\psi(r) =( \psi_{\rm b} - \psi_{\rm c} )\frac{\log(r/r_{\rm c})}{\log (r_{\rm b}/r_{\rm c})}  +  \psi_{\rm c}
\end{equation}
Eq. (\ref{director_sol}) is independent of the elastic constant $K$.
Eq. (\ref{director_sol}) indicates constant ratios between  $r_i$ and $r'_i$ in the intensity profile.
The difference $\psi(r_i) -  \psi(r'_i)$ may be $\pi/4$ or $-\pi/4$.
Thus, we have 
\begin{equation}
\log (r_i/r'_i)=\log(r_{\rm b}/r_{\rm c})(\psi(r_i) -  \psi(r'_i))/(\psi_{\rm b}-\psi_{\rm c})
\end{equation}
where the right-hand term is a constant.
In the results shown in Figs.\ref{dynamical} and \ref{multi_ring}, the properties of a ring pattern, 
such as the size and extinction time, may depend on $r_{\rm b}$ and $\psi_{\rm b}$ originated in other disclinations.
To  control the size of the pattern, it is essential to control the distortion in the director field far from the laser spot.  

Eq. (\ref{director_sol}) gives the director field on the pattern. 
The director on the pattern rotates along the radial direction.
In the texture of a multiple-ring pattern, the director on a dark ring is perpendicular to that of the adjacent rings.
Thus, a multiple-ring pattern is found as multiple circular walls (MCW) \cite{deGennes}.

\begin{figure}
\includegraphics[scale=0.8]{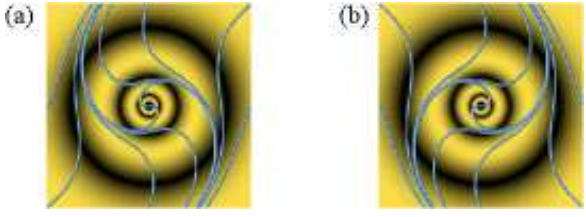}
\caption{
(color online).  Existence of a chiral pair in MCW. 
The intensity of both MCWs are based on Eq. (\ref{director_sol}).
The director fields (blue lines) are calculated from Eq. (\ref{director_sol}).
(a) Clockwise MCW ($\psi_{\rm b}$=-1.75$\pi$). 
(b) Counterclockwise MCW ($\psi_{\rm b}$=1.75$\pi$).
The parameters $r_{\rm b}/r_{\rm c}$=32.0, $\psi_{\rm c}$=0 are the same in the two cases.
In the experiments, chirality is determined by the trajectory of scanning.}
\label{simulated_ring}
\end{figure}

Figure \ref{simulated_ring} shows the  existence of a chiral pair in MCWs calculated from Eq. (\ref{director_sol}). 
There are clockwise (Fig. \ref{simulated_ring} (a)) and counterclockwise (Fig. \ref{simulated_ring} (b)) MCWs, 
where the intensity profiles of Fig. \ref{simulated_ring} (a) and (b) are identical.
In addition, the elastic energies of MCW in Fig. \ref{simulated_ring} (a) and (b) are also degenerated.
In the experiments, we observed the chirality of MCWs by using the analyzer rotation technique \cite{Tabe03}.
This chirality in MCWs is controlled by choosing the proper trajectory of laser scanning. 

There have been several reports of target patterns with many rings in a thin film of SmC LC \cite{Tabe03,Cladis85,Cladis95,Stannarius,uto}.
If we compare our results to those reports, the C-director of the SmC phase plays the role of the director of the nematic.
Especially, an expression similar to Eq. (\ref{director_sol}) was previously obtained for 
the azimuth angle of C-director \cite{Cladis85}. 

\section{Conclusion}
We have reported a novel method for generating MCWs, 
which are a stable  director field in nematic LC, through the use of proper laser scanning.
We have shown that the chirality of MCWs can be controlled by choosing a suitable trajectory of laser scanning.

\ack	
This work was supported by 
Technology of Japan and by a Sasakawa Scientific Research Grant (No. 19-643) from The Japan Science Society,
Grant-in-aid for young researchers from Kyoto University Venture Business Laboratory (KU-VBL)
and a Grant-in-Aid for Scientific  
Research on Priority Areas (No. 17076007) from the Ministry of Education, Culture, Sports and Science.  

\bibliography{bibliography.bib}
\end{document}